\documentclass[fleqn,usenatbib]{mnras}

\usepackage[T1]{fontenc}
\usepackage{ae,aecompl}

\usepackage{graphicx}   
\usepackage{amsmath}    
\usepackage{amssymb}    

\def\l{\lambda}
\def\m{\mu}

\def\1{{\bf 1}}

\def\tt{\therefore}
\def\bra{\langle}
\def\ket{\rangle}

\newcommand{\ol}\overline
\newcommand{\ul}\underline
\newcommand{\ti}\tilde
\newcommand{\wt}\widetilde
\newcommand{\wh}\widehat

\newcommand{\ft}{\footnote}

\newcommand{\be}{\begin{equation}}
\newcommand{\ee}{\end{equation}}
\newcommand{\bl}{\begin{eqnarray}&}
\newcommand{\el}{&\end{eqnarray}}
\newcommand{\bq}{\begin{eqnarray}}
\newcommand{\eq}{\end{eqnarray}}
\newcommand{\bqsn}{\begin{eqnarray*}}
\newcommand{\eqsn}{\end{eqnarray*}}

\title[Luminosity Function in QSOs by MaxEnt]{The Luminosity Function of Quasars by the Principle of Maximum Entropy}
\author[A. Andrei et al.]{
Alexandre Andrei $^{1,2,3}$ \thanks{E-mail: oat1@on.br }, Bruno
Coelho $^{4}$ \thanks{E-mail: brunodfcoelho@gmail.com }, 
Leandro Guedes $^{5}$ \thanks{E-mail: leandrolsguedes@planeta.rio}
and Alexandre Lyra$^{1}$\thanks{E-mail: alexandr@astro.ufrj.br}
\\
$^{1}$Universidade Federal do Rio de Janeiro, Observat\'orio do Valongo, Ladeira Pedro Antonio, 43, Rio de Janeiro, \\ RJ, CEP 20080-090, Brazil\\
$^{2}$ Observat\'orio Nacional, MCTIC, Rua Gal. Jos\'e Cristino 77, Rio de Janeiro, CEP 20921-400, Brazil \\
$^{3}$ SYRTE, Observatoire de Paris, 61 Avenue de l'Observatoire, 75014 Paris, France  \\
$^{4}$ Instituto de Telecomunica\c{c}\~oes, Campus Universit\'ario de Santiago, 3810-193 Aveiro, Portugal\\
$^{5}$Planetarium Foundation of the City of Rio de Janeiro, Rua Vice-Governador R\'ubens Berardo 100 - G\'avea, Rio de Janeiro - \\ RJ, CEP 22451-070, Brazil\\
}

\date{Accepted XXX. Received YYY; in original form ZZZ}

\pubyear{2018}

\begin{document}
\label{firstpage}
\pagerange{\pageref{firstpage}--\pageref{lastpage}}    \maketitle
\begin{abstract}
We propose a different way to obtain the distribution of the luminosity function of quasars by using the Principle of Maximum Entropy. The input data comes from the SDSS-DR3 quasars counts, extending up to redshift 5 and limited from apparent magnitude $i=15$ to 19.1 at $z\lesssim3$ to $i=20.2$ for $z\gtrsim3$. Using only few initial data points, the Principle allows us to estimate probabilities and hence that luminosity curve. 
We carry out statistical tests to evaluate our results. The resulting luminosity function compares well to earlier determinations. And our results remain consistent either when the amount or choice of sampled sources is unbiasedly altered. 
Besides this we estimate the distribution of the luminosity function for redshifts in which there is only observational data in the vicinity.
\end{abstract}



\section{Introduction}

The quasar luminosity function gives a measure for the bidimensional distribution of quasars in luminosity and redshift. Fundamentally it indicates that the universe is not in a stationary state. As consequence it requires the due interpretation before using quasars to determine cosmological parameters, but at the same time it informs about the evolution of quasars themselves, and the changing content of the space intervening between distant quasars and the observer. The function usually describing the quasar luminosity function, as a function of redshift and absolute luminosity, basically starts from the modulus distance formulae and incorporates several corrections, to accommodate line emission, the expanding universe scale of distance, the intrinsic dependency of quasar light emission on wavelength, terms of self and media absorption, etc. The result is an empirical description, which exponents and coefficients are adjusted to each sample examined. It is interesting thus to build an independent function, able to describe the quasar luminosity function in a simpler form and from different physical principles. Although by necessity also incorporating the astrophysical and cosmological assumptions, an alternative, simpler form for the quasar luminosity function can be derived from the statistical mechanics  methods. 

The concept of entropy, since Clausius, 
became part of thermodynamics. In addition, 
it also became part of statistical mechanics. 
The study of systems in equilibrium and out of equilibrium is closely related to the notions of 
entropy as well as its production.
There is a vast bibliography about it with warm discussions. 
We can cite three important related principles: 
Ziegler's {\it maximum entropy
production principle} (e. g. \citet{Ziegler1}, \citet{Ziegler}, 
see also \citet{Dewar05});
Prigogine's {\it minimum entropy production principle}   
(\citet{Prigogine-67}, \citet{Prigogine-78},
\citet{Kondepudi}) 
\ft{This principle is the subject of a specific work  in
 \citet{Jaynes1980}.}; 
and the Maximum Entropy Principle (MaxEnt) (\citet{Jaynes-1}).    
This paper employs the last one. 

 Derivations of the first two principles from \mbox{MaxEnt} can be found in literature, as seen in  \citet{Martyushev}.
In that review the authors make a very interesting description of the 
MaxEnt focusing on the production of entropy. Other authors emphasize that Jaynes's
MaxEnt formulation of statistical mechanics provides a theoretical
basis for Maximum Entropy Production Principle (\citet{Dewar-Maritan}).
The applications of the MaxEnt are many. 
We'll see below related issues and discuss how they are connected to the focus of our treatment, that aims to find the distribution 
of the luminosity function of quasars.

Despite this vast reach there are authors who restrict
the  MaxEnt applications (see eg. \citet{Synthese-Shimony-1}, and references therein). 

Some of these critiques were addressed by Jaynes himself (\citet{Jaynes Responde-1}). In this paper Jaynes also provides a
fairly complete description of MaxEnt from its roots to its
implications. On the other hand, we can not fail to mention that there are also papers written
exactly in defense of Jaynes's Principle as in
\citet{Tikochinsky et al-1}, stating with: ``{\it The only consistent
algorithm is one that leads to the distribution of Maximum entropy
subject to constraints given.}''. There are other papers with very
interesting critiques that bring out points for and against MaxEnt and provide quite compelling references on the subject, like in the 
Appendix A of \citet{MNRAS-Pontzen et al-2013}, 
where the authors sketch Jaynes's reasoning, ``{\it that the maximization 
of entropy subject to certain constraints is equivalent to testing whether 
these constraints encapsulate
later the physics of the situation...}'', and the
use of the method to derive the phase space distribution of a
virialized dark matter halo.  

In addition, there are 
several other areas in physics and astrophysics where it can be applied. Some examples are, in spectral analysis
(\citet{Ables1}), where ``{\it the method produces superior spectral
representations when compared with more traditional methods...}''.
as well as a powerful technique of image reconstruction
(\citet{Skilling-Bryan}), in the same paper other applications 
of MaxEnt in astronomy can be found. In \citet{Gull-Daniell} MaxEnt
is applied in radio and X-ray astronomy. 
It is interesting that the method is also applied in X-ray
tomographic image reconstruction and restoration (\citet{Ali
Mohammad1}). In the case of astrophysics and cosmology, we see
papers where the dark energy equation of state $w(z)$ is
reconstructed using the MaxEnt (\citet{Zunckel and Trotta}).

In gravitation, with the confirmation in 2016 of the existence of the gravitational 
waves predicted by A. Einstein, the study of the black 
holes assumes still greater importance. The earliest detections 
were precisely on collisions of black-holes (\citet{Abbott et al.}).
The traditional second law of thermodynamics was modified into a generalized second law for the study of black holes (\citet{Bekenstein-1974}).  The   
Jaynes's method of maximum entropy was also used by Bekenstein 
to determine the probability distribution for a system containing a Kerr black hole 
(\citet{Bekenstein-1975}).

This paper presents a new approach to find the distribution of probabilities of the luminosity function using the MaxEnt technique. Even with some criticisms like those
cited above, we believe that the MaxEnt is extremely useful to be applied when we have partial information about a certain system. So this principle allows us to know accurate 
probabilities (see formula (\ref{eq_prob_princ}))  from a small data set. 
Although the number of known quasars is constantly increasing, 
to get perhaps to a million known objects in the next decade, 
small subsamples are useful and had not been yet
designed by lack of elements. On one hand, the quasars zoo is 
also growing, different types of active galaxies conceivably exhibiting 
luminosity functions peculiar by a certain degree. On the other hand, 
the capability of mapping in detail particular thin slices of the 
universe in redshift is long sought, nonetheless to better define 
the complex form of the luminosity function. Finally, it is important 
to be able to drawn different samples of a large dataset for sanity 
check control. Is this paper we will explore such capability of the MaxEnt description
of the luminosity distribution.

Since quasars discovery (\citet{Schmidt}, \citet{Matthews}) their energy
output and magnitude have been object of much observation and
increasingly complex theories. Conversely,
 that information became much used for studies as surrounding host galaxies,
 gravitational lenses, in situ and
intergalactic absorption, up to the cosmological scale of
distances in an expanding  universe. The so-called
 luminosity function is all important 
 to make sense of
such extraordinary energy output and to those astrophysical
quantities from it derived.  
The evolution of the quasar luminosity function with redshift is an 
important observational tool, that allows us to put constraints on the formation 
and growth history of supermassive black holes, and 
their co-evolution with host galaxies. 
It also give us a measure of the contribution of quasars 
in the cosmological reionization of the Universe. 
For all these the study of the quasar luminosity function 
has received the attention of 
several works (eg. \citet{Richards et al-1}, 
\citet{Masters}, \citet{Ross}, \citet{Manti}).

So, among some successful applications of MaxEnt in astrophysics, we are going now to explore a new one, in the study of the quasar luminosity function.

The Sloan Digital Sky Survey (SDSS) 
\ft{$http://www.sdss.org/$}
provided observations of quasars in different redshifts, 
being responsible for the identification of the vast majority 
of the known Quasi-Stellar Objects (QSOs; \citet{Paris}).  
However there 
are observational limitations, to the effect  that one can ask: what would be the quasar 
distribution on each redshift slice if we could consider unobserved magnitudes?
These observational limitations must be taken into account when computing 
the quasar luminosity function, however this does not constitute 
the aim of this work. So we are going to use already corrected 
counts of QSOs computed by \citet{Richards et al-1} in their 
study of the quasar luminosity function.  
We show here that the MaxEnt can provide a good distribution 
of probabilities for the luminosity 
function from few values of a limited sample in each redshift.   


The luminosity function provides the density distribution of 
classes of objects, per unit volume and assuming a statistically 
complete sample. In the case of quasars this indicates more 
or less probable scenarios for their formation and evolution, 
as well as their relationship with the host galaxy. 
Quasars have been found out by several projects, chiefly 
the SDSS, relying on different strategies to single 
them out from the more numerous contaminants of other
celestial bodies. The ESA cornerstone mission Gaia 
combines the recognition of such known quasars, with 
micro arcsecond determination of proper motions over 
five years,  therefore providing direct means to cleanse 
away the intruding false positives, as nearby red dwarfs. On top of it,  Gaia 
will use a neural network strategy leading to 
autonomous recognition of quasars. Combined to 
the all sky repeated sweeping of objects up to 
near-red twentieth-second magnitude, it will 
produce an unprecedent complete sample of quasars. Therefore to establish an alternative, independent, 
and physics robust method of tracing the quasar luminosity 
function affords a strong way of checking upon and getting 
feedback from the usual Schechter based determination.
In short, the motivations for these studies are threefold, 
an independent study of the luminosity function on the
quasar population in the SDSS DR3, the development of an 
independent tool for determining the luminosity function based 
on maximum entropy physics, and it is a comparative assessment 
on a limited sample with views to application on the all sky, 
statistically complete sample of quasars in final Gaia catalogue.

Elseways the definition of the quasars 
luminosity function have so far been done using a 
modified template of the Schechter exponential for 
galaxies. Such a description although well adapted 
to the somehow simpler quasar case, 
since it is in practice free from the surface brightness issue,  
limits the reliability of the astrophysical and cosmological interpretation of the luminosity function. 
To mention a few, it is known that the shape and turnover 
of the luminosity function would favor either models for 
the growth of the super massive black hole from mergers 
or by inflow and host galaxy instabilities. The bright 
end of the luminosity function can favor intrinsic properties 
about which time black holes are increasing in mass 
rapidly whereas the faintest end would indicate about 
the length of time quasars spend at relatively low accretion rates.


The remainder of this work is organized as follows. 
In section \ref{sec_jaynes_approach} we will briefly review the MaxEnt method. 
With this we establish our main formula, the equation (\ref{eq_prob_princ}), which defines from MaxEnt the probability of the luminosity function. 
In section \ref{MaxEnt_Luminosity} we will summarize our technique to determine 
the luminosity function of quasars, and we show the  details of 
how the Lagrange multipliers were calculated for the studied cases, 
in addition  we will show the comparison between our result by MaxEnt and 
the Schechter's based \citet{Richards et al-1}  one.


In  section \ref{Graficos-testes} 
we will describe the statistical tests we use.  
In section \ref{new section1} from few observational data in particular redshifts, 
we will make a prediction of the PDF (Probability Density Function)  for in between  redshifts,
that is, we will estimate the distribution of the luminosity function. 
Finally, a summary discussion 
and conclusions are presented in the last section.
\section{The Jaynes Approach to Maximum Entropy Principle}\label{sec_jaynes_approach} 
We can sum up the Maximum Entropy Principle as we shall see in the
sequel\footnote{Here we will follow \citet{Jaynes-1}.}.
As there is a vast bibliography regarding this principle, 
we will only make a brief account. 

Initially we assume that a quantity $x$ can have the discrete values $x_i (i
=1,2, ..., n)$, but we do not know the corresponding probabilities
$p_i$. All we have is the expectation value of the function
$f(x)$,
\bq \langle f(x) \rangle = \sum_{i=1}^n p_i f(x_i)
\label{eq_vlr_esperado} \eq
Based on this information, how can we obtain the expectation value
of another function of the system  $g(x)$?   Jaynes responds to this apparently
insoluble question. The given information is insufficient to
determine the probabilities $p_i$. The equation
(\ref{eq_vlr_esperado}) and the normalization condition
\bq  \sum p_i = 1 \label{eq_som_prob_ig_1} \eq
would have to be supplemented by $(n-2)$ more conditions before
$g(x)$ could be found.

In order to find a solution to this problem, Jaynes's method uses
the following expression for entropy
\bq H(p_1, p_2, ..., p_n) = - k  \sum_i p_i \ln p_i \;\; ,
\label{eq_entropia} \eq
where $k$ is a positive constant. Since $H$ is just the expression
for entropy as found in statistical mechanics, it will be called
the ``entropy of the probability distribution $p_i$''. 
The entropy $H$, given in (\ref{eq_entropia}) is maximized subject to the
constraints (\ref{eq_vlr_esperado}) and (\ref{eq_som_prob_ig_1}). 

In order to achieve a final expression for the probability of $x_i$, 
we use the method of Lagrangian multipliers, usually noted by $\l$ and $\m$, 
where $\l$  is associated with the normalization equation, i.e. 
the equation (\ref{eq_som_prob_ig_1}) and $\m$ is associated with 
the equation of the expectation value (\ref{eq_vlr_esperado}). 
With this methodology we obtain the probability
\bq p_i = e^{- \l - \m f(x_i)} \label{eq_prob_princ}  \; . \eq
This formula gives an important expression, which can be associated to the function of 
the luminosity distribution of the objects to which we wish to estimate 
the distribution, and the method used in its determination is called 
the Maximum Entropy Principle. See the complete development from data
to Lagrange multipliers at Appendix \ref{apendice_multiplicadores}.
\section{The Luminosity Function of QSO(s) From MaxEnt } 
\label{MaxEnt_Luminosity}
To summarize what will be done next, from MaxEnt we will 
determine the distribution of the luminosity function of the 
quasars in a certain redshift $z_k$, by using the probability distribution (\ref{eq_prob_princ}). 
Notice that the strong energy released by quasars make possible 
to observe them from the nearby Universe at least up to redshifts greater than 7 (eg. \citet{banados}). This large range of distance, hence an evolving luminosity function, 
allowed us to inspect how consistent are the predictions from MaxEnt, 
and compare the results against those originally derived from the same observed data, 
used here as control result.

In the MaxEnt methodology, for the consistency of the principle, 
the strongest symmetry that we could have ``a priori'' would be the uniform 
distribution, but this is not the case. We know that if we have a single constraint, 
that is associated with normalization $\sum_{i=1}^n p_i  = 1$ , we get exactly for 
$n=N \rightarrow p_i = 1 / N $  
or the uniform distribution. The other constraint, associated with equation 
(\ref{eq_vlr_esperado}), 
breaks this symmetry. Let us also remember that as it is well placed in \citet{Caticha-2004} : 
``{\it The method of maximum entropy (ME) is designed for
updating from a prior probability distribution to a posterior
distribution when the information to be processed takes the
form of a constraint ... }''. Then, we assume that we can extract a certain expected value obtained through some 
luminosity values provided by the system observations, 
which obviously have the uniformity between all values broken. 
These values are randomly chosen, and under these conditions we will apply MaxEnt with their two  constraints: 
(\ref{eq_vlr_esperado}) and (\ref{eq_som_prob_ig_1}). This is the central point of the methodology, 
namely that from just some values a strong estimate of the 
luminosity function of the distribution of all values in this 
redshift can be made\footnote{One interesting question posed by Jaynes is: ``generating paradoxes 
in the case of continuously variable random quantities, 
since intuitive notions of ``equally possible'' are 
altered by a change of variables'' (Jaynes (1957)p.622).}.

For the present quasar luminosity function derivation by MaxEnt, 
we have tested different sets from the whole of the initial data, 
seeing in every case great accordance between the Luminosity curve 
from MaxEnt and the control result. In order to analyze 
the most realistic scenario, the one for which the sample 
is small and, thus, not necessarily containing a perfect
representation of the data population, we choose to analyze 
here the results from random initial data. We have picked up just 
three luminosities in each redshift as initial data. 

The starting point of using MaxEnt is the calculation of  
$\m$ and $\l$ from the equations (\ref{eq_vlr_esperado}) 
and (\ref{eq_som_prob_ig_1}) (see details in Appendix \ref{apendice_multiplicadores}). 
From a certain redshift, the mean value to be used in the 
Lagrange multiplier method is calculated from three 
luminosities randomly chosen, to each of which is assigned 
the corrected number of quasars in that luminosity bin after 
applying the  selection function of 
\citet{Richards et al-1}, Table
6, p. 2782. Those values will be used to calculate the weighted mean luminosity $\bra L_z \ket $, which
is the value to be used in the equation (\ref{eq_vlr_esperado}). 
The other Lagrange multiplier comes
from the normalization of the probability, or, $\sum_{i=1}^n p_i  = 1$.

Errors have been calculated using a bootstrap method. 
In each case, three random luminosities were 
drawn 200 times and the mean value used to find a different $\l$ and $\m$ 
that, applied to original data, gave us a different set of points. 
The extreme values stand as the upper and lower limits of 
the error bars to the results from the principle. 
Likewise the errors on the control result were 
calculated using probabilities from bootstrap draws.

Verifying our assumptions, the calculated probabilities 
by MaxEnt and the ones of the control results show similar behavior.

For each redshift the complete table leading to the control 
result is in Appendix \ref{apendice_RichardTable}, 
Table \ref{tabela-do-richard}., and the three ones 
randomly chosen in each redshift are on the lines 
indicated in bold at the first column.

The conversion from calibrated magnitudes to luminosities was done using the following relation 
\bq
L&=&10^{\frac{-(M_i+48,6)}{2,5}} 4\pi \left(3,0857 \times 10^{19}\right)^2 \mbox{,}
\label{conversao_1}
\eq
where $L$ (in ergs s$^{-1}$ Hz$^{-1}$) is the luminosity and $M_i$ the magnitude. 

The curves obtained for each redshift are shown in Figure 1. We can see clearly that a correspondence is found at the sampled redshifts, within the error bars, between the MaxEnt results and those for the control.

\begin{figure*}
\hspace{-2.1cm}
\includegraphics[scale=0.6]{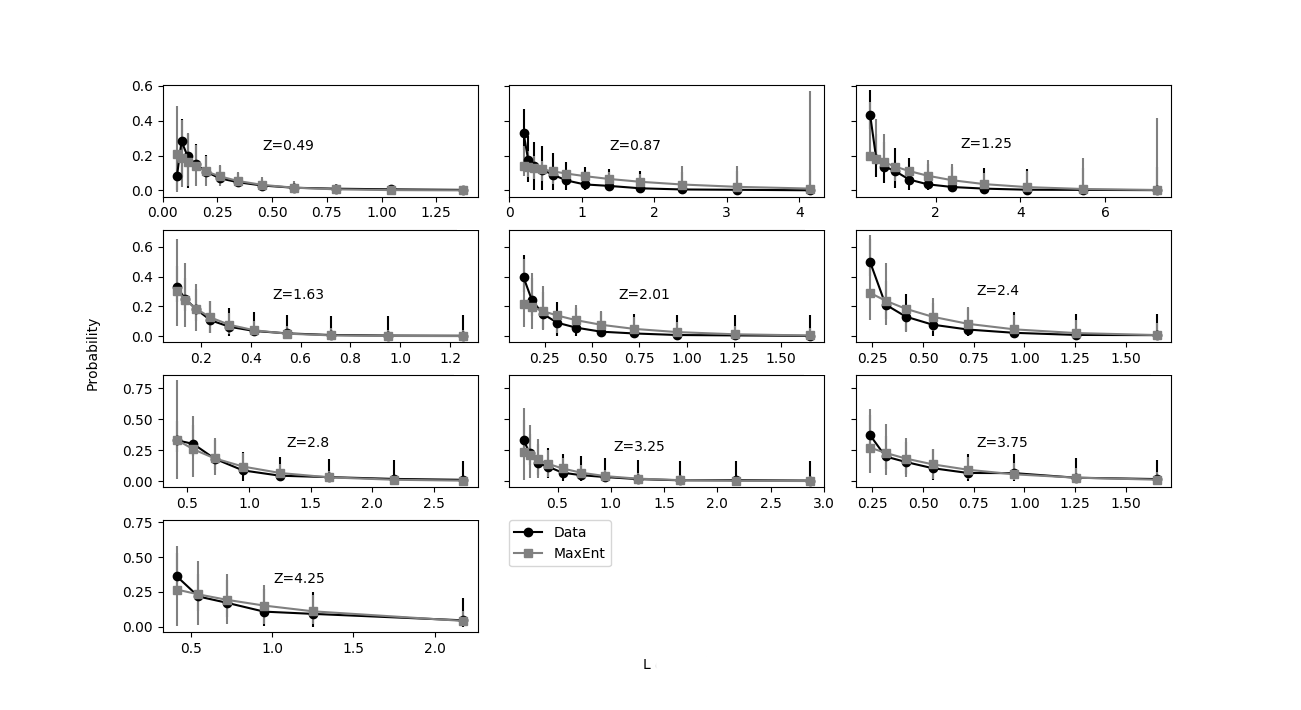}

\caption{Comparison between probabilities calculated from the MaxEnt method and from the control results. Plots show luminosities on the horizontal axis and probabilities on the vertical ones. Values at horizontal axis should be multiplied by $10^{-31}$ (ergs s$^{-1}$ Hz$^{-1}$)  at Z=0.49, 0.87, 1.25 and by $10^{-32}$ (ergs s$^{-1}$ Hz$^{-1}$) at the others.}
\label{figcase2}
\end{figure*}

\section{Statistical Tests} 
\label{Graficos-testes}
As discussed in the previous paragraphs, and detailed in Appendix \ref{apendice_multiplicadores}, 
the MaxEnt approach, from robust yet simple physical principles and 
computational algorithms, delivers 
 a statistical probability distribution of the luminosity function which is cosmologically plausible, vis a vis the literature on the subject. The magnitude and redshift data used 
for that is taken from the SDSS project. It is natural thus that 
the outcomes from the luminosity function here obtained shall be 
compared with those from the SDSS analysis. 

At the start of the current application of the MaxEnt 
principle to derive the quasars luminosity function, 
several approaches were used. Choosing by hand representative data, 
choosing data from quartiles of the distribution, and picking up 
the extreme and mean values. The outcomes were always concurrent 
(they are available under request), what served as sanity check, 
as well as gave us ground to adopt the random draws finally used. 
The plots in Figure \ref{figcase2} are compelling to show the agreement between 
the two luminosity function statistical probability distributions. 
Such agreement can be quantified. Table \ref{spearman} shows results of  
the statistical tests comparing the two distributions of probabilities, 
the one from MaxEnt and the one from the control results, at each redshift. 

As indicated in section \ref{MaxEnt_Luminosity} a minimal number of points were randomly drawn from the data. Using only these few data points, MaxEnt can provide us an estimate 
luminosity function to be compared with the luminosity function obtained 
from the control results. The results are compared by verifying 
the mutual correlation. The Spearman's correlation test is used 
because of the small number of chosen points, as well to not
assume their normal distribution. The $\rho$ is quite close to unity. Notice however that although the luminosity function is best represented as an exponential progression, the pair of points of the two compared distributions are not 
necessarily so, thus we have also used the F-test  and the Student's T-Test because these are nonparametric tests. 

Since the shape of the curves is obviously similar but not 
the error bars, while the number of points is small, 
the F-test for variances is advisable. 
The table of the F- distribution indicates 
that the null hypothesis (no difference) must 
be accepted to a large degree of statistical 
certainty, with two exceptions, out of the limit redshift. 
Those exceptions lie at $z=0.87$ and $z=1.25$, for which the null 
hypothesis certainty is mediocre. In both cases, 
that befalls upon the large error bars seeing at 
the one brightest luminosities. The F-Test without 
those points give us results 1.64 and 0.92 respectively, 
that take us back to a null hypothesis scenario. 

On views of the outturn of the correlation and variance 
tests pointing to the agreement of the MaxEnt and control
results, the two-samples Student's T-test is next justified. 
On this one, as Table \ref{spearman} shows, in all cases -- even for the 
troublesome redsfifts as detected in the previous tests -- the null-hypothesis on the means can not be rejected for 
usual statistical standards.  

Table \ref{spearman} brings the three statistics for the distributions. 
On Table \ref{testes-erros}, instead, the same statistical tests are applied 
to compare their error budgets. Notice, at start, 
that the error bars are asymmetrical, and therefore up 
and down pairs are formed. The correlations are
poor, though they undeniably exist. On the other hand, 
the F-test and T-test for the errors 
show the MaxEnt method and the control results faring quite alike also in this respect. 
We thus can further conclude for the 
independence of the methods, but similar efficiency.


\begin{table}
\centering
\caption{Statistical tests 
comparing the MaxEnt and control results luminosity curves.}
\label{spearman}
\begin{tabular}{lllllll}
     &                     &                  &              &                 &              &\\
     &\multicolumn{2}{c|}{ \bf{Spearman}}         &\bf{F-Test}  & \multicolumn{2}{c|}{ \bf{Student-t}}               \\
     & $\;\;$ \bf{$\rho$}  &\bf{P-Value}  &              & \bf{T-status}   & \bf{P-value} &\\
0.49 & 0.93                & $1.17 \times 10^{-5}$   & 1.32         &   $\ll 10^{-99}$             & 1            &\\   
0.87 & 1                   & $\ll 10^{-99}$                & 4.40         &8.88 $\times 10^{-16}$ & 0.99         &\\
1.25 & 1                   & $\ll 10^{-99}$                & 3.26         &   $\ll 10^{-99}$             & 1            &\\   
1.63 & 0.99                & $6.65 \times 10^{-64}$  & 1.10         &   $\ll 10^{-99}$             & 1            &\\   
2.01 & 0.99                & $6.65 \times 10^{-64}$  & 2.75         &   $\ll 10^{-99}$             & 1            &\\   
2.4  & 1                   & $\ll 10^{-99}$                & 2.53         & $1.99 \times 10^{-16}$ & 0.99         &\\
2.8  & 1                   & $\ll 10^{-99}$                & 1.12         & $2.21 \times 10^{-16}$ & 0.99         &\\ 
3.25 & 0.99                & $3.76 \times 10^{-9}$   & 1.48         & $3.35 \times 10^{-16}$ & 0.99         &\\  
3.75 & 1                   & $\ll 10^{-99}$                & 1.50         &   $\ll 10^{-99}$             & 1            &\\   
4.25 & 1                   & $\ll 10^{-99}$                & 1.87         &   $\ll 10^{-99}$             & 1            &\\   
\end{tabular}
\end{table}
\begin{table}
\centering
\caption{Statistical tests 
comparing the error budgets over the MaxEnt
and control results luminosity curves.}
\label{testes-erros}
\begin{tabular}{lllllll}
     &                     &                  &              &                 &              &\\
     &\multicolumn{2}{c|}{ \bf{Spearman}}         &\bf{F-Test}  & \multicolumn{2}{c|}{ \bf{Student-t}}               \\
     & $\;\;$ \bf{$\rho$}  &\bf{P-Value}  &              & \bf{T-status}   & \bf{P-value} &\\
0.49 & 0.77 & $1.02\times10^{-5}$ & 0.70 & 1.48  & 0.15 &\\   
0.87 & 0.48 & 0.02           & 0.19 & 0.71 & 0.48 &\\
1.25 & 0.63 & 0.00           & 0.22 & 0.11  & 0.91 &\\   
1.63 & 0.60 & 0.00           & 0.29 & 0.44  & 0.67 &\\   
2.01 & 0.53 & 0.02           & 0.48 & 0.03 & 0.97 &\\   
2.4  & 0.46 & 0.07           & 0.29 & 0.50  & 0.62 &\\
2.8  & 0.60 & 0.01           & 0.16 & 0.55  & 0.58 &\\ 
3.25 & 0.56 & 0.01           & 0.41 & 0.45 & 0.65 &\\  
3.75 & 0.65 & 0.01           & 0.41 & 0.45  & 0.65 &\\   
4.25 & 0.58 & 0.05           & 0.24 & 1.49  & 0.15 &\\   
\end{tabular}
\end{table}



\section{Estimation of the luminosity function for other redshifts }
\label{new section1}
\begin{figure*}
\includegraphics[scale=.5]{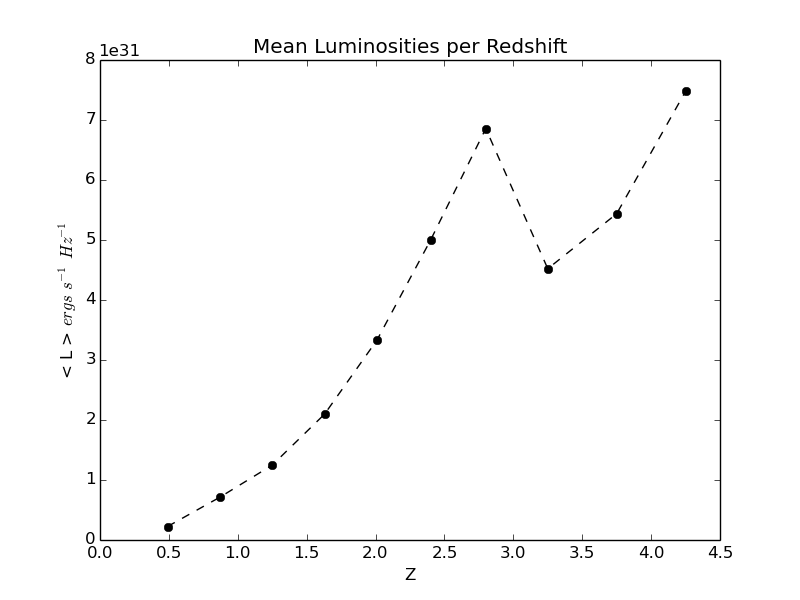}
\caption{Mean luminosity in bins of redshift from $z=0.5$ to $z=4.25$ derived from SDSS DR3 as in \citet{Richards et al-1}, not taking into account any corrections for the Malmquist bias.}

\label{graf-med-lum}
\end{figure*}

\begin{figure*}
\centering
  \includegraphics[scale=0.55]{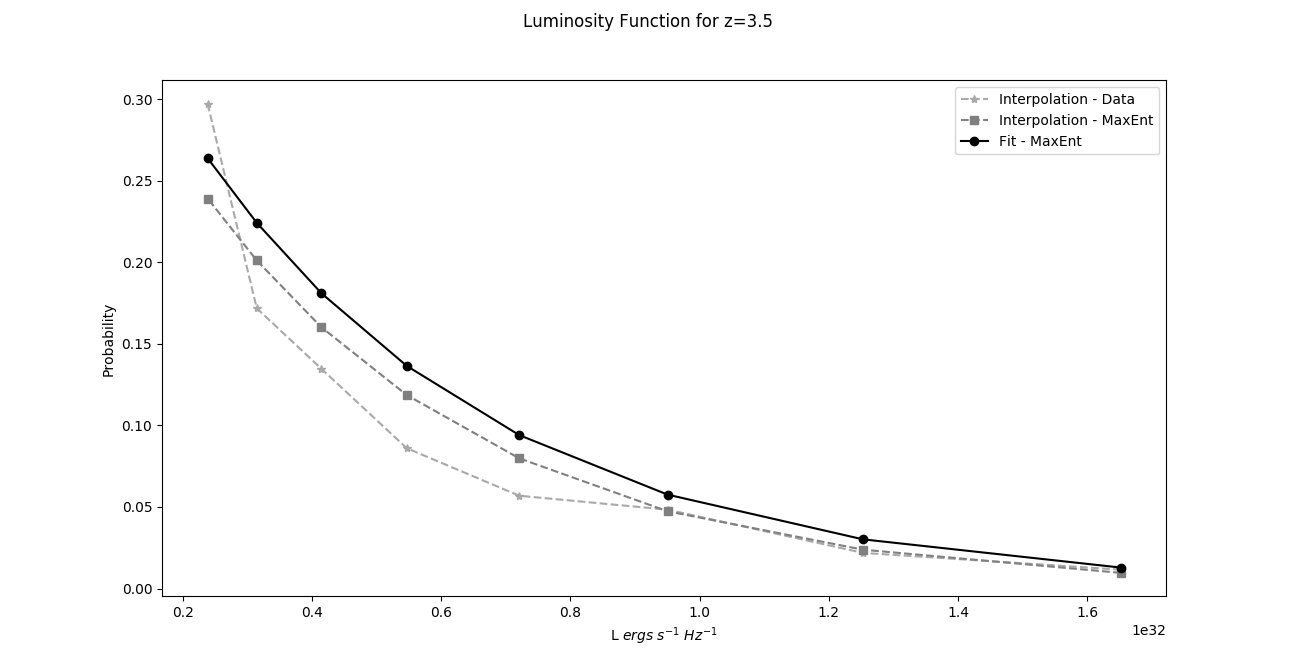}
\caption{Luminosity function obtained from MaxEnt using the fitting curve to z=3.5.}
\label{graf-aj-z-35}
\end{figure*}

In this section we use the MaxEnt luminosity function presented in this paper to investigate the outcomes for a redshift in which we suppose that data exist only in its vicinity.

For this simulation, the redshift $z=3.5$ is chosen. As shown in Figure 2 at this redshift the luminosity $L$ seems to increase again after a drop between $z=2.75$ and $z=3.25$, at the same time there are enough input data and good results for the neighbor redshifts. From those the  mean value $\langle L_{z=3.5} \rangle$ is interpolated,  and next we  will  obtain  by  MaxEnt  the  distribution of $L$ for the redshift 3.5.

In practice, we start from  the  same  set  of data used before, from Richards et al (2006), plotting all the available redshifts with respective mean luminosities. Then the curve of best fitting to the observational data is obtained, and from this fitting curve we associate a mean luminosity to the redshift aimed at. Next, in order to procure the Lagrange multipliers $\mu$ and $\lambda$ a set of observed luminosities is demanded. Those were picked up at random from the luminosities actually present for the neighbor redshifts.  

The point now is to verify whether using this quite arbitrary choice the MaxEnt formulation is capable to issue a credible luminosity function. We thus compare the MaxEnt formulation results to a direct interpolation of the control results and of the MaxEnt results themselves (both depicted on Figure 1). Figure 3 shows these three results. It is seen that the MaxEnt formulation based on neighbor data gives a result comparable to the direct interpolation results, but at the same time it delivers a smoother curve.

This  type of  situation occurs  frequently  in  astrophysics,  and  MaxEnt demonstrates here to be a very useful tool to estimate values, what later can be tested later as more data becomes available.


\section {Conclusions}

The quasar luminosity function is intended as a measure of the actual distribution of quasars in luminosity and redshift. For that observational, astrophysical, and cosmological restricting factors  must be accounted for and often different surveys must be combined, before a complete population is inferred. That satisfied, most quasar luminosity functions available in the literature are represented either by a double power-law regimen or by a modified Schechter function. The adjustments are semi-empirical, having as usual parameters a normalization factor, a break magnitude, a reference redshift, and bright and faint ends slopes.

By contrast, the MaxEnt method, on top of being quite simple to handle, offers three strong features. First it represents a physically distinct approach, thus bringing the known benefits of different bias, limitations, and systematics. Secondly because it is purely statistical, it depends of less astrophysical and cosmological assumptions, in special the key ones break magnitude and reference redshift. Thirdly, a hallmark of MaxEnt is to deliver trustful conclusions from small samples. This last quality is particularly suited to deal with limited dedicated surveys, as well as to piece off portions of the luminosity function without further requirements to the mathematical representation of the function itself. By the same token it is suited to try out luminosity functions for putative new classes of quasars and their location, either within large clusters or relatively isolated.

In this pioneer derivation we took the SDSS DR3 quasar population, and the normalization made by \citet{Richards et al-1} there in. The luminosity functions and corresponding curves were used here as control results. The comparisons hold very well, being practically immaterial whether the whole luminosity population or samples as small as three random elements were used.

As Jaynes has stated, that MaxEnt is the generalization of the Principle of Insufficient Reason. In our case, we show that little information of the system (quasars luminosities) gave us consistent results. In so it is an effective way of practical generalization. As a result, the Lagrange multipliers behaved in a stable manner, enabling to use bootstrapping for determination  of errors. The aspect of updating the knowledge when of the outcome of a much larger data set, as expected from Gaia, is foreseen to be coherently accommodated, as well as to investigate piecemeal the luminosity function.

\section*{Acknowledgments} 
A. L. thanks the colleagues at the Valongo Observatory, 
H. M. Boechat Roberty and M. Assafin, for suggestions in the beginning of this work.
A. A. thanks CPNp Grant Bolsa de 
Produtividade em Pesquisa 306775/2018.
B. C. acknowledges support from the Advanced EU Network of E-infrastructures for Astronomy with SKA (AENEAS), funded by the European Commission Framework Programme Horizon 2020 RIA under grant agreement n. 731016 and from the ENGAGE SKA RI, grant POCI-01-0145-FEDER-022217, funded by COMPETE 2020 and FCT, Portugal. We must also thank 
the anonymous referee for valuable suggestions and comments.
\appendix
\section{Table A1} \label{apendice_RichardTable}
Table \ref{tabela-do-richard}   
was obtained from \citet{Richards et al-1}, with the addition 
of the probability required to our objective and derived from their data, 
plus the probability we obtained for the comparison.

\begin{table}
\centering
\caption{The Redshift, Luminosities and Probability}

\label{tabela-do-richard}

\begin{tabular}{llll}
\bf{Z}& \bf{L ($\times 10^{30}$)}   & \bf{Prob}    & \bf{Prob}\\
    & (\bf{erg/s/Hz})        &              & \bf{MaxEnt} \\
0.49  &  13.70  &  $3.32\times 10^{-3}$ &  $3.84 \times 10^{-4}$\\
0.49  &  10.42  &  $6.336\times 10^{-3}$&  $1.89\times 10^{-3}$\\
0.49  &  7.90  &  $1.00\times 10^{-2}$  &  $6.36\times 10^{-3}$\\
\bf{0.49} &  5.60 &  $1.55\times 10^{-2}$&  $1.59\times 10^{-2}$\\
0.49  &  4.55  &  $2.83\times 10^{-2}$  &  $3.20\times 10^{-2}$\\
\bf{0.49}  & 3.45  &  $4.70\times 10^{-2}$ & $5.42\times 10^{-2}$\\
0.49  & 2.62  &  $6.95\times 10^{-2}$  &  $8.09\times 10^{-2}$\\
\bf{0.49}  &  1.99  &  $1.08\times 10^{-1}$  & $1.10\times 10^{-1}$\\
0.49  &  1.51  &  $1.52\times 10^{-1}$  &  $1.38\times 10^{-1}$\\
0.49  &  1.14  &  $1.96\times 10^{-1}$  &  $1.65\times 10^{-1}$\\
0.49  &  0.87 &  $2.83\times 10^{-1}$  &  $1.88\times 10^{-1}$\\
0.49  &  0.66  &  $8.16\times 10^{-2}$  &  $2.08\times 10^{-1}$\\
0.87  &  41.50  &  $9.74\times 10^{-4}$  &  $1.07\times 10^{-2}$\\
0.87  &  31.50  &  $3.70\times 10^{-3}$  &  $2.07\times 10^{-2}$\\
\bf{0.87}  &  23.90  &  $5.61\times 10^{-3}$  &  $3.40\times 10^{-2}$\\
0.87  &  18.10  &  $1.23\times 10^{-2}$  &  $4.96\times 10^{-2}$\\
0.87  &  13.70  &  $2.62\times 10^{-2}$  &  $6.61\times 10^{-2}$\\
0.87  &  10.40  &  $3.46\times 10^{-2}$  &  $8.21\times 10^{-2}$\\
0.87  &  7.91  &  $5.74\times 10^{-2}$  &  $9.69\times 10^{-2}$\\
0.87  &  5.60  &  $9.09\times 10^{-2}$  &  $1.10\times 10^{-1}$\\
\bf{0.87}  &  4.55  &  $1.2\times 10^{-1}$  &  $1.21\times 10^{-1}$\\
0.87  &  3.45  &  $1.419\times 10^{-1}$  &  $1.30\times 10^{-1}$\\
0.87  &  2.62  &  $1.77\times 10^{-1}$  &  $1.37\times 10^{-1}$\\
\bf{0.87}  &  1.99  &  $3.30\times 10^{-1}$  &  $1.43\times 10^{-1}$\\
1.25  &  72.10  &  $1.69\times 10^{-3}$  &  $2.79\times 10^{-3}$\\
1.25  &  54.70  &  $3.29\times 10^{-3}$  &  $8.38\times 10^{-3}$\\
\bf{1.25}  &  41.50  &  $4.14\times 10^{-3}$  &  $1.93\times 10^{-2}$\\
1.25  &  31.50  &  $1.04\times 10^{-2}$  &  $3.63\times 10^{-2}$\\
1.25  &  23.90  &  $1.98\times 10^{-2}$  &  $5.87\times 10^{-2}$\\
1.25  &  18.10  &  $3.52\times 10^{-2}$  &  $8.45\times 10^{-2}$\\
1.25  &  13.70  &  $6.27\times 10^{-2}$  &  $1.11\times 10^{-1}$\\
\bf{1.25}  &  10.40  &  $1.11\times 10^{-1}$  &  $1.37\times 10^{-1}$\\
1.25  &  7.90  &  $1.37\times 10^{-1}$  &  $1.61\times 10^{-1}$\\
\bf{1.25}  &  5.60  &  $1.807\times 10^{-1}$  &  $1.82\times 10^{-1}$\\
1.25  &  4.55  &  $4.33\times 10^{-1}$  &  $1.99\times 10^{-1}$\\
1.63  &  125.00  &  $2.57\times 10^{-3}$  &  $1.55\times 10^{-4}$\\
1.63  &  95.00  &  $3.79\times 10^{-3}$  &  $1.14\times 10^{-3}$\\
1.63  &  72.10  &  $7.57\times 10^{-3}$  &  $5.20\times 10^{-3}$\\
1.63  &  54.70  &  $1.86\times 10^{-2}$  &  $1.64\times 10^{-2}$\\
\bf{1.63}  &  41.50  &  $3.46\times 10^{-2}$  &  $3.93\times 10^{-2}$\\
\bf{1.63}  &  31.50  &  $6.30\times 10^{-2}$  &  $7.61\times 10^{-2}$\\
1.63  &  23.90  &  $1.07\times 10^{-1}$  &  $1.26\times 10^{-1}$\\
1.63  &  18.10  &  $1.82\times 10^{-1}$  &  $1.84\times 10^{-1}$\\
\bf{1.63}  &  13.70  &  $2.51\times 10^{-1}$  &  $2.46\times 10^{-1}$\\
1.63  &  10.40  &  $3.31\times 10^{-1}$  &  $3.06\times 10^{-1}$\\
2.01  &  165.00  &  $1.65\times 10^{-3}$  &  $4.48\times 10^{-3}$\\
\bf{2.01}  &  125.00  &  $4.87\times 10^{-3}$  &  $1.25\times 10^{-2}$\\
2.01  &  95.00  &  $7.54\times 10^{-3}$  &  $2.71\times 10^{-2}$\\
2.01  &  72.00  &  $1.81\times 10^{-2}$  &  $4.89\times 10^{-2}$\\
2.01  &  54.70  &  $3.00\times 10^{-2}$  &  $7.65\times 10^{-2}$\\
\bf{2.01}  &  41.50  &  $5.72\times 10^{-2}$  &  $1.07\times 10^{-1}$\\
2.01  &  31.50  &  $9.06\times 10^{-2}$  &  $1.39\times 10^{-1}$\\
2.01  &  23.90  &  $1.50\times 10^{-1}$  &  $1.69\times 10^{-1}$\\
\bf{2.01}  &  18.10  &  $2.44\times 10^{-1}$  &  $1.96\times 10^{-1}$\\
2.01  &  13.70  &  $3.96\times 10^{-1}$  &  $2.19\times 10^{-1}$\\
2.4  &  165.00  &  $6.57\times 10^{-3}$  &  $7.35\times 10^{-3}$\\
\bf{2.4}  &  125.00  &  $8.08\times 10^{-3}$  &  $2.07\times 10^{-2}$\\
2.4  &  95.00  &  $2.22\times 10^{-2}$  &  $4.56\times 10^{-2}$\\
2.4  &  72.00  &  $4.44\times 10^{-2}$  &  $8.28\times 10^{-2}$\\
\bf{2.4}  &  54.70  &  $7.71\times 10^{-2}$  &  $1.30\times 10^{-1}$\\
2.4  &  41.50  &  $1.31\times 10^{-1}$  &  $1.84\times 10^{-1}$\\
\bf{2.4}  &  31.50  &  $2.12\times 10^{-1}$  &  $2.39\times 10^{-1}$\\
2.4  &  23.90  &  $4.98\times 10^{-1}$  &  $2.91\times 10^{-1}$\\
\end{tabular}
\end{table}

\begin{table}
\centering
\contcaption{}
\begin{tabular}{llll}
\bf{Z}& \bf{L ($\times 10^{30}$)}   & \bf{Prob}    & \bf{Prob}\\
    & (\bf{erg/s/Hz})        &              & \bf{MaxEnt} \\

    2.8  &  274.00  &  $1.04\times 10^{-2}$  &  $3.64\times 10^{-3}$\\
2.8  &  218.00  &  $1.77\times 10^{-2}$  &  $1.09\times 10^{-2}$\\
\bf{2.8}  &  165.00  &  $3.21\times 10^{-2}$  &  $3.01\times 10^{-2}$\\
2.8  &  125.00  &  $4.34\times 10^{-2}$  &  $6.53\times 10^{-2}$\\
\bf{2.8}  &  95.00  &  $8.65\times 10^{-2}$  &  $1.17\times 10^{-1}$\\
\bf{2.8}  &  72.10  &  $1.81\times 10^{-1}$  &  $1.83\times 10^{-1}$\\
2.8  &  54.70  &  $3.00\times 10^{-1}$  &  $2.57\times 10^{-1}$\\
2.8  &  41.50  &  $3.29\times 10^{-1}$  &  $3.32\times 10^{-1}$\\
3.25  &  287.00  &  $4.09\times 10^{-3}$  &  $4.64\times 10^{-4}$\\
3.25  &  218.00  &  $7.11\times 10^{-3}$  &  $2.32\times 10^{-3}$\\
\bf{3.25}  &  165.00  &  $6.63\times 10^{-3}$  &  $7.85\times 10^{-3}$\\
\bf{3.25}  &  125.00  &  $1.63\times 10^{-2}$  &  $2.00\times 10^{-2}$\\
3.25  &  95.00  &  $3.25\times 10^{-2}$  &  $3.99\times 10^{-2}$\\
3.25  &  72.10  &  $4.80\times 10^{-2}$  &  $6.80\times 10^{-2}$\\
3.25  &  54.70  &  $6.79\times 10^{-2}$  &  $1.02\times 10^{-1}$\\
\bf{3.25}  & 41.50  &  $1.15\times 10^{-1}$  &  $1.38\times 10^{-1}$\\
3.25  &  31.50  &  $1.45\times 10^{-1}$  &  $1.75\times 10^{-1}$\\
3.25  &  23.90  &  $2.25\times 10^{-1}$  &  $2.08\times 10^{-1}$\\
3.25  &  18.10  &  $3.32\times 10^{-1}$  &  $2.38\times 10^{-1}$\\
3.75  &  165.00  &  $1.65\times 10^{-2}$  &  $1.14\times 10^{-2}$\\
3.75  &  125.00  &  $2.74\times 10^{-2}$  &  $2.78\times 10^{-2}$\\
3.75  &  95.00  &  $6.42\times 10^{-2}$  &  $5.48\times 10^{-2}$\\
\bf{3.75}  &  72.10  &  $6.57\times 10^{-2}$  &  $9.16\times 10^{-2}$\\
\bf{3.75}  &  54.70  &  $1.04\times 10^{-1}$  &  $1.35\times 10^{-1}$\\
3.75  &  41.49  &  $1.54\times 10^{-1}$  &  $1.82\times 10^{-1}$\\
\bf{3.75}  &  31.50  &  $1.98\times 10^{-1}$  &  $2.28\times 10^{-1}$\\
3.75  &  23.90  &  $3.69\times 10^{-1}$  &  $2.70\times 10^{-1}$\\
\bf{4.25}  &  218.00  &  $4.65\times 10^{-2}$  &  $4.22\times 10^{-2}$\\
4.25  &  125.00  &  $9.28\times 10^{-2}$  &  $1.11\times 10^{-1}$\\
\bf{4.25}  &  95.00  &  $1.09\times 10^{-1}$  &  $1.53\times 10^{-1}$\\
4.25  &  72.10  &  $1.73\times 10^{-1}$  &  $1.94\times 10^{-1}$\\
4.25  &  54.70  &  $2.18\times 10^{-1}$  &  $2.33\times 10^{-1}$\\
\bf{4.25}  &  41.50  &  $3.61\times 10^{-1}$  &  $2.67\times 10^{-1}$\\
\end{tabular}
\end{table}

\section{Lagrange Multipliers Method: Data, Constraints and
Computation} \label{apendice_multiplicadores}
To develop the fundamentals of MaxEnt, consider the following set of data
\begin{center}
\begin{tabular}{ll}
Object$_1$     & $A_1$ \\
Object$_2$     & $A_2$\\
Object$_3$     & $A_3$\\
...          & {\it ...}     \\
Object$_n$     & $A_n$\\
\end{tabular}
\end{center}
\noindent where each $\textnormal{Object}_i$ is a quasar and $A_i$
its respective luminosity, with $i=(1,2,3,...,n)$. From now on we adapt Jaynes's notation to our work. Thus, we will
call the luminosities by $ A_i$, and their mean value by $\langle A
\rangle$. Each $A_i$ has a probability $p_i$ to occur and we get from the data an average value $\langle A \rangle$ that can be obtained from arithmetic mean, weighted average, or from a more
accurate form, using expression
(\ref{eq_vlr_esperado}). This expression may be rewritten as
\bq \langle A \rangle = \sum\limits_{i=1}^n A_ip_i \;\; , 
\label{media_pela_probabilidade_melhorada}
\eq
where at one redshift $z_k$, the index $n$ varies in the sum of $i = 1,...,n$ 
only on selected objects, that is, only in those three chosen 
luminosities in this redshift.
Considering that the data set contains all possible values to
occur, we have the bond condition that the summation of all
probabilities must be equal to $1$, see (\ref{eq_som_prob_ig_1}), or
\bq
 \sum\limits_{i=1}^{n} p_i = 1
 \label{somatorio_de_p_igual_a_um}
\eq
The two Lagrange multipliers $\m$ and $\l$ are
associated to these two equations respectively,
(\ref{media_pela_probabilidade_melhorada}) and
(\ref{somatorio_de_p_igual_a_um}). Then next they will be placed into a new form of the above equations.

From \ref{media_pela_probabilidade_melhorada} we have 
\bq
  \m \left[ \sum\limits_{i=1}^{n} A_ip_i - \langle A \rangle \right]=0 \; \; \mbox{,}
 \label{lamb}
\eq
and from \ref{somatorio_de_p_igual_a_um}
\bq
 \l \left[ \sum\limits_{i=1}^{n} p_i - 1 \right]=0 \mbox{.}
 \label{m}
\eq
According to Jaynes, the  method consists in the determination of the
distribution function, $p_i$,  by maximizing the so-called
informational entropy
\bqsn
 H\equiv H(p_i,p_2,...p_n)=-K\sum\limits_{i=1}^{n} p_i \ln p_i \; \mbox{,}
\eqsn
\noindent this can be done by the standard method using the
additional conditions (\ref{media_pela_probabilidade_melhorada}) and
(\ref{somatorio_de_p_igual_a_um}) and the Lagrange multipliers
$\l$ and $\m$.
The maximization procedure leads to the following result
\bq
 p_i= e^{-\m A_i -\l}  \; .
 \label{probabilidae_melhorada}
\eq
The two equations that we have to adjust to compute are obtained
by taking (\ref{probabilidae_melhorada}) into the equations of
constraints (\ref{media_pela_probabilidade_melhorada}) and
(\ref{somatorio_de_p_igual_a_um}), so we obtain the equations
\bq e^{-\l}\sum\limits_{i=1}^{n}A_i e^{-\m A_i} &=&
\langle A \rangle\label{primeira_equacao_para_integrar} \\
e^{-\l} \sum\limits_{i=1}^{n} e^{-\m A_i} &=& 1
\label{segunda_equacao_para_integrar} \; .  \eq
The equation (\ref{primeira_equacao_para_integrar}) inform us
that
\bq e^{\l}=\frac{\sum\limits_{i=1}^{n}A_i e^{-\m
A_i}}{\langle A \rangle} \; . \label{01/12/2017:12:02}\eq
Taking (\ref{01/12/2017:12:02}) into equation (\ref{segunda_equacao_para_integrar}) we
obtain an equation  in $\m$ to be solved:
\bq \langle A \rangle \frac{\sum\limits_{i=1}^{n}e^{-\m A_i}}{\sum\limits_{i=1}^{n}A_i e^{-\m A_i}} = 1
\label{expressao_final_lambda_1} .\eq
To find $\l$, the obtained values of $\m$  are taken into (\ref{01/12/2017:12:02}). That is, the sequence of procedures to find $\m$ from equation (\ref{expressao_final_lambda_1}) and substitute it into the equation
(\ref{01/12/2017:12:02}) to find $\l$.

With both Lagrange Multipliers found, we can get by MaxEnt the
resulting probability (\ref{probabilidae_melhorada}) and the
average value (\ref{media_pela_probabilidade_melhorada}) for each
redshift.




\end{document}